\documentclass[12pt]{article}
\usepackage{amsmath}
\usepackage{color}
\usepackage{amssymb}
\usepackage{bm}

\makeatletter\@addtoreset{equation}{section}
\makeatother
\allowdisplaybreaks[1]
\begin{document}
\begin{titlepage}

\begin{flushright}
\phantom{preprint no.}
\end{flushright}
\vspace{0.5cm}
\begin{center}
{\Large \bf
New Chandrasekhar transformation in\\
\vspace{2mm}
Kerr spacetime
}
\lineskip .75em
\vskip0.5cm
{\large Hiroaki Nakajima${}^{1}$ and Wenbin Lin${}^{1,\,2,\,*}$}
\vskip 2.5em
${}^{1}$ {\normalsize\it School of Mathematics and Physics, University of South China, \\ Hengyang, 421001, China\\
}
\vskip 1.0em
${}^{2}$ {\normalsize\it School of Physical Science and Technology, Southwest Jiaotong University, \\ Chengdu, 610031, China\\}
\vskip 1.0em
${}^{*}$ {\normalsize\it Email: lwb@usc.edu.cn\\}
\vskip 1.0em
\vskip 3.0em
\end{center}
\begin{abstract}
We construct a new type of Chandrasekhar transformation
in Kerr spacetime using the different tortoise coordinate, which is useful for exact analysis to study Teukolsky equation with arbitrary frequency.
We also give the interpretation of our transformation using the formalism of the quantum Seiberg-Witten geometry.
\end{abstract}
\end{titlepage}

\section{Introduction}
The direct observation of gravitational waves by LIGO and Virgo \cite{Abbott:2016blz} has opened the new era of cosmology.
The binary system such as the black hole merger is a good object to observe the gravitational waves,
where the theoretical calculation is also possible.
One of the method for the calculation of the gravitational waves radiated from the binary system
is the black hole perturbation theory \cite{Mino:1997bx},
which is in particular useful for the case of the extreme mass-ratio inspiral (EMRI).
In this formalism, we have to solve the linearized Einstein equation in the black hole spacetime.
Fortunately, it was found that the separation of the variables is possible for particular gauges.
For the Schwarzschild spacetime, the two equations for the radial variable have been obtained: the Regge-Wheeler equation
\cite{Regge:1957td} and the (radial) Teukolsky equation \cite{Teukolsky:1973ha}.
Those equations are derived from the same linearized Einstein equation
but with the different gauges, and hence it is expected that there is a relation between the Regge-Wheeler and
the Teukolsky equation, originated by the gauge transformation. The explicit one-to-one correspondence was
found by Chandrasekhar \cite{Chandrasekhar:1975}, called the Chandrasekhar transformation.
For the Kerr spacetime, it is not known how to obtain the Regge-Wheeler-type equation directly
from the linearized Einstein equation, and only the Teukolsky equation is obtained.
Hence some Regge-Wheeler-type equations \cite{Detweiler:1977gy,Chandrasekhar:1976zz,Sasaki:1981sx}
are proposed using the Chandrasekhar(-like) transformation from the Teukolsky equation.
The Chandrasekhar transformation is also known as an example of the Darboux transformation
\cite{Darboux,Glampedakis:2017rar,Yurov:2018ynn}.

Mathematically, the Regge-Wheeler and the Teukolsky equations belong to the confluent Heun's equation (CHE) \cite{Heun},
which has two regular singularities and one irregular singularity. CHE also has the so-called accessory parameter,
which cannot be reproduced from the local leading behavior of the solution. Due to the existence of
and the accessory parameter, solving the equation globally is very difficult and so far some local behavior of the solutions
around the regular singularities are mainly studied. Moreover, the solution which is regular at the origin is recently
implemented in Mathematica. For the global solution,
the expression as the series of the hypergeometric functions \cite{Heun}, corresponding to
the solution of the Teukolsky equation with the low-frequency expansion, have just been established
\cite{Mano:1996vt,Casals:2021ugr}.

Recently,
CHE have also been found as the differential equation associated with
the quantum Seiberg-Witten geometry \cite{Seiberg:1994rs,Seiberg:1994aj,NS} in supersymmetric gauge theories.
Moreover, due to AGT (Alday-Gaiotto-Tachikawa) correspondence \cite{Alday:2009aq,Gaiotto:2009ma},
the same equation can also be regarded as the BPZ (Belavin-Polyakov-Zamolodchikov) equation
in two-dimensional conformal field theory \cite{Bonelli:2021uvf}.
These correspondence is helpful to study the solution to the Regge-Wheeler and the Teukolsky equations
beyond the low-frequency approximation.
It is also found that the Chandrasekhar(-like) transformation in the Schwarzschild spacetime can be
interpreted as the exchange of the mass parameters \cite{Aminov:2020yma,Hatsuda:2020iql}.
Before that it has already been known that this exchange of the parameters is regarded as a particular integral transform
\cite{Whiting:1988vc,Andersson:2016epf}.
Performing the integral transform needs to know the global behaviour of the function.
On the other hand, since the Chandrasekhar(-like) transformation just consists of the function itself and its derivative,
one can easily find the local behaviour (around the regular singularities, in particular) of the transformed function.
In this paper, we will consider a new Chandrasekhar(-like) transformation in the Kerr spacetime,
which can be interpreted as the transform of the mass parameters the quantum Seiberg-Witten geometry. Moreover, it would also help to study the problem with arbitrary frequency~\footnote{If the frequency is high enough, the analysis using the geometrical optics is available.}.

The reminder of this work is organized as follows: in section 2, we review the Chandrasekhar transformation,
which is extended from the original work by introducing a constant parameter for later convenience.
In section 3, we consider the new Chandrasekhar(-like) transformation in the Kerr spacetime.
In section 4, we interpret our new transformation as the change of the parameters in CHE
using the formalism of the quantum Seiberg-Witten geometry.
Section 5 is devoted to summary and discussion.

\section{Chandrasekhar transformation}
We first review about the general procedure of the Chandrasekhar transformation \cite{Chandrasekhar:1975}.
Let the function $X(x)$ satisfy the differential equation
\begin{equation}
\left[\Lambda_{-}\Lambda_{+}-V_{X}(x)\right]X=0,
\label{V}
\end{equation}
where the differential operators $\Lambda_{\pm}$ are defined by
\begin{equation}
\Lambda_{\pm}=\frac{d}{dx}\pm ip(x).
\label{lambda}
\end{equation}
Then the Chandrasekhar transformation \cite{Chandrasekhar:1975} is given by
\begin{equation}
Y=FX+G\Lambda_{+}X,
\label{C-transf1}
\end{equation}
where $F$ is taken to be
\begin{equation}
F=\alpha^{-1} V_{X},
\label{F}
\end{equation}
with constant $\alpha$, and
$G$ is some function of $x$ to be determined later.
Acting $\Lambda_{-}$ to the both hand sides of \eqref{C-transf1} gives
\begin{equation}
\Lambda_{-}Y=AX+B\Lambda_{+}X,
\label{C-transf2}
\end{equation}
where $A$ and $B$ are defined by
\begin{align}
A&=F'-2ipF+GV_{X},
\label{C-transf3A}
\\
B&=F+G',
\label{C-transf3B}
\end{align}
and the prime denotes the derivative with respect to $x$.
Again, acting $\Lambda_{+}$ to the both hand sides of \eqref{C-transf2} and eliminating $X$ gives
\begin{equation}
\Lambda_{+}\Lambda_{-}Y-\frac{A'}{A}\Lambda_{-}Y-\alpha BY
=\left(A+B'+2ipB-\alpha BG-\frac{B}{A}A'\right)
\Lambda_{+}X.\label{Yeq1}
\end{equation}
In order to make the above be the closed equation for $Y$, we require
that the right hand side of \eqref{Yeq1} should vanish, namely
\begin{equation}
G=\alpha^{-1}\left(2ip+\frac{A}{B}-\frac{A'}{A}+\frac{B'}{B}\right).
\label{Weq1}
\end{equation}
Then $Y$ satisfies
\begin{equation}
\Lambda_{+}\Lambda_{-}Y-\frac{A'}{A}\Lambda_{-}Y-\alpha BY=0.
\label{Yeq2}
\end{equation}
Finally, by multiplication transformation $Y=H\tilde{Y}$, we will obtain the desired form of the differential equation.
For example, when we choose $H=A^{\frac{1}{2}}$, the differential equation for $\tilde{Y}$ becomes
\begin{equation}
\left[\Lambda_{+}\Lambda_{-}-V_{Y}\right]\tilde{Y}=0, \quad
V_{Y}=B-\frac{H''}{H}-2ip\frac{H'}{H}+2\left(\frac{H'}{H}\right)^{2},
\end{equation}
which has similar form as \eqref{V}, but the order of $\Lambda_{+}$ and $\Lambda_{-}$ is reversed.

Let us recapitulate the procedure of the transformation for the case of the Schwarzschild spacetime.
In this background, $X$ is assumed to satisfy the Regge-Wheeler equation \cite{Regge:1957td} in the frequency domain
\begin{gather}
\left[\left(\frac{r-2M}{r}\frac{d}{dr}\right)^{2}+\omega^{2}
-V_{\mathrm{RW}}(r)\right]X=0.
\label{RW}
\\
V_{\mathrm{RW}}(r)=
\left(1-\frac{2M}{r}\right)\left[\frac{l(l+1)}{r^{2}}
-\frac{6M}{r^{3}}\right],
\label{RWpot}
\end{gather}
on the other hand the (radial) Teulkolsky equation \cite{Teukolsky:1973ha} with spin $s=-2$ in the frequency domain is
\begin{gather}
\left[(r^{2}-2Mr)\frac{d^{2}}{dr^{2}}-2(r-M)\frac{d}{dr}+U_{\mathrm{T}}(r)
\right]R=0, \label{T}
\\
U_{\mathrm{T}}(r)=
\left(1-\frac{2M}{r}\right)^{-1}\left[(\omega r)^{2}-4i\omega (r-3M)\right]
-(l-1)(l+2).
\end{gather}
Here $r$ is the standard radial coordinate, $\omega$ is the frequency of the gravitational waves,
$M$ is the mass of the black hole,
and $l$ denotes the multipole which takes the value $l=2,3,\ldots$
By introducing the dimensionless coordinate $z=\frac{r}{2M}$,
the Regge-Wheeler equation \eqref{RW} and
the Teukolsky equation \eqref{T} can be rewritten as
\begin{gather}
\left[\left(\frac{z-1}{z}\frac{d}{dz}\right)^{2}+\epsilon^{2}
-\left(1-\frac{1}{z}\right)\left(\frac{l^{2}+l}{z^{2}}-\frac{3}{z^{3}}\right)\right]X=0~.
\label{RWz}
\\
z(z-1)\frac{d^2 R}{dz^2}-(2z-1)\frac{dR}{dz}
+\left[\frac{z}{z-1}\left(\epsilon^2 z^{2}-2i\epsilon(2z-3)\right)-(l^2+l-2)
\right]R=0~,
\label{Tz}
\end{gather}
where $\epsilon$ is the dimensionless frequency parameter defined by
\begin{equation}
\epsilon=2M\omega.
\end{equation}
We will use the tortoise coordinate
\begin{equation}
z^{\ast}=z+\ln (z-1),
\label{tortoise1}
\end{equation}
as our variable $x$. The differential operators $\Lambda_{\pm}$ \eqref{lambda} are taken to be
\begin{equation}
\Lambda_{\pm}=\frac{d}{dz^{\ast}}\pm i\epsilon=\left(1-\frac{1}{z}\right)\frac{d}{dz}\pm i\epsilon.
\end{equation}
Using  $\Lambda_{\pm}$, the Regge-Wheeler equation \eqref{RWz} can be rewritten in the form of
\eqref{V} with
\begin{equation}
V_{X}=\left(1-\frac{1}{z}\right)\left(\frac{l^{2}+l}{z^{2}}-\frac{3}{z^{3}}\right).
\end{equation}
In \cite{Chandrasekhar:1975}, Chandrasekhar chose $\alpha=1$ and then the function $F$ is just $V_{X}$, and chose $G$ in \eqref{C-transf1} as
\begin{equation}
G=\frac{2z-3}{z^{2}}+2i\epsilon~.
\label{Wspin2}
\end{equation}
Then $A$ and $B$ are computed from \eqref{C-transf3A} and \eqref{C-transf3B} as
\begin{equation}
A=\frac{3}{z^{4}}\left(1-\frac{1}{z}\right)^{2}~, \quad
B=\left(1-\frac{1}{z}\right)\left(\frac{l^{2}+l-2}{z^{2}}+\frac{3}{z^{3}}\right)~,
\end{equation}
One can confirm that \eqref{Weq1} is indeed satisfied.
The differential equation \eqref{Yeq2} for
$Y$ becomes
\begin{equation}
\Lambda_{+}\Lambda_{-}Y+\frac{2(2z-3)}{z^{2}}\Lambda_{-}Y
-\left(1-\frac{1}{z}\right)\left(\frac{l^{2}+l-2}{z^{2}}+\frac{3}{z^{3}}\right)Y=0~.
\label{Yeq3}
\end{equation}
Finally \eqref{Yeq3} can be rewritten into the Teukolsky equation \eqref{Tz} by
the multiplication transformation $Y=z^{-3}R$.
Thus the Chandrasekhar transformation is
\begin{align}
R&=z^{3}\left[V_{X}X+\left(\frac{2z-3}{z^{2}}+2i\epsilon\right)\Lambda_{+}X\right]
\notag\\
&=z^{2}f\Lambda_{+}f^{-1}\Lambda_{+}zX~,
\label{C-transf}
\end{align}
where $f=1-z^{-1}$.

\section{New transformation in Kerr spacetime}
Now we consider a similar Chandrasekhar-like transformation for the Teukolsky equation in Kerr spacetime.
We use the conventional Boyer-Lindquist radial coordinate $r$. The outer and the inner
horizons in this coordinate are located at
\begin{equation}
r=r_{\pm}=M\pm \sqrt{M^2 - a^2},
\end{equation}
where $a$ is the Kerr parameter. The limit $a \to 0$ corresponds to the Schwarzschild spacetime.
The Teukolsky equation in Kerr spacetime with the spin $s=-2$ in the frequency domain \cite{Teukolsky:1973ha} is given by
\begin{gather}
\Delta\frac{d^2 R}{dr^2}-2(r-M)\frac{dR}{dr}+U_{\mathrm{T}}R=0,
\\
U_{\mathrm{T}}=\frac{K^2+4i(r-M)K}{\Delta}-8i\omega r-\lambda,
\end{gather}
where $\lambda$ is the eigenvalue of the equation, which approaches to $l^2+l-2$
in $a \to 0$ limit. $\Delta$ and $K$ are defined by
\begin{align}
\Delta&=r^2 -2Mr+a^2=(r-r_{+})(r-r_{-}),
\\
K&=(r^2 + a^2)\omega -am.
\end{align}
Here $m$ can take the integer values with $-l \le m \le l$.
In order to make the structure of the equation simpler, we introduce the dimensionless
coordinate $z$ by%
\footnote{We assume that Kerr black hole is non-extremal. i.e. $a<M$.
The extremal case ($a=M$) should be considered separately. }
\begin{equation}
z=\frac{r-r_{-}}{r_{+}-r_{-}}.
\label{z}
\end{equation}
By this transformation the regular and the irregular singularities $r=r_{-}, r_{+}, \infty$ of the Teukolsky equation
are mapped into $z=0,1,\infty$, respectively.
The Teukolsky equation in terms of the $z$-coordinate becomes
\begin{equation}
z(z-1)\frac{d^2 R}{dz^2}-(2z-1)\frac{dR}{dz}
+\left[\frac{k^2+2i(2z-1)k}{z(z-1)}-4i\tilde{k}-\lambda
\right]R=0,
\label{Tzk}
\end{equation}
where $k$ and $\tilde{k}$ are given by
\begin{gather}
k=\mathcal{A}\epsilon z^2 +\mathcal{B}\epsilon z+\mathcal{C},\quad
\tilde{k}=\frac{dk}{dz}=2\mathcal{A}\epsilon z +\mathcal{B}\epsilon,
\\
\mathcal{A}=\frac{r_{+}-r_{-}}{2M}, \quad
\mathcal{B}=\frac{r_{-}}{M}, \quad
\mathcal{C}=\frac{r_{-}\epsilon -am}{r_{+}-r_{-}}.
\end{gather}
Note that under the limit $a \to 0$, the above quantities behave as
\begin{equation}
k \to \epsilon z^{2}, \quad \tilde{k}\to 2\epsilon z, \quad \mathcal{A}\to 1,\quad \mathcal{B},\ \mathcal{C}\to 0.
\end{equation}

Since it is not known how to obtain the Regge-Wheeler-type equation in Kerr spacetime directly
from the linearized Einstein equation, we here consider the Chandrasekhar-like transformation from the Teukolsky equation.
As the independent variable $x$ we use the coordinate $z^{\ast}$ \eqref{tortoise1} defined from \eqref{z}.
Note that in Kerr spacetime, $z^{\ast}$ is different from the conventional tortoise coordinate $z^{\ast\ast}$
defined by
\begin{align}
z^{\ast\ast}&=z+\frac{2Mr_{+}}{(r_{+}-r_{-})^2}\ln (z-1) - \frac{2Mr_{-}}{(r_{+}-r_{-})^2}\ln z
\notag\\
 &=z+\frac{1+\mathcal{A}}{2\mathcal{A}^{2}}\ln (z-1) - \frac{1-\mathcal{A}}{2\mathcal{A}^{2}}\ln z~.
\label{tortoise2}
\end{align}
In literature \cite{Detweiler:1977gy,Chandrasekhar:1976zz,Sasaki:1981sx},
$z^{\ast\ast}$ is used as the independent variable, since $d/dz^{\ast\ast}$ is the Killing vector field
for the gravitational wave radiation. 
However, the differential equation using $z^{\ast\ast}$ 
has the apparent singularities at $z=(-1+\mathcal{A}\pm i\sqrt{1-\mathcal{A}^2})/2\mathcal{A}$ (corresponding to $r=\pm ia$),
which makes the analysis (in particular the discussion in the next section) complicated.
Here we use the coordinate $z^{\ast}$ since the apparent singularities do not appear
and also the resulting equation is simpler.
We take the differential operators $\Lambda_{\pm}$ as
\begin{equation}
\Lambda_{\pm}=\frac{d}{dz^{\ast}}\pm i\frac{k}{z^2}
=\left(1-\frac{1}{z}\right)\frac{d}{dz}\pm
i\left(\mathcal{A}\epsilon+\frac{\mathcal{B}\epsilon}{z}+\frac{\mathcal{C}}{z^2}\right).
\end{equation}
Then the Teukolsky equation \eqref{Tz} is rewritten as
\begin{equation}
\Lambda_{+}\Lambda_{-}R-\frac{2}{z}\Lambda_{-}R-\frac{z-1}{z^{3}}(3i\tilde{k}+\lambda)R=0~.
\end{equation}
By the multiplication transformation $Y=z^{-3}R$, the differential equation for $Y$ becomes
\begin{equation}
\Lambda_{+}\Lambda_{-}Y+\frac{2(2z-3)}{z^{2}}\Lambda_{-}Y
-\left(1-\frac{1}{z}\right)\left(\frac{\lambda -3i\mathcal{B}\epsilon}{z^{2}}+\frac{3-6i\mathcal{C}}{z^{3}}\right)Y=0~.
\end{equation}
From the above, $A$ and $B$ can be read off as
\begin{equation}
A=\frac{\alpha^{-1} c_{0}}{z^{4}}\left(1-\frac{1}{z}\right)^{2}~, \quad
B=\alpha^{-1}\left(1-\frac{1}{z}\right)
\left(\frac{\lambda -3i\mathcal{B}\epsilon}{z^{2}}+\frac{3-6i\mathcal{C}}{z^{3}}\right)~.
\end{equation}
where $c_{0}$ is constant. $G$ is computed from \eqref{Weq1} as
\begin{equation}
\alpha G=\frac{2ik}{z^{2}}+\frac{2z-3}{z^{2}}+(c_{0}-3+6i\mathcal{C})\frac{z-1}{z^{3}}
\left(\lambda -3i\mathcal{B}\epsilon+\frac{3-6i\mathcal{C}}{z}\right)^{-1}.
\end{equation}
By choosing $c_{0}=3-6i\mathcal{C}$, the above can be simplified as
\begin{equation}
\alpha G=\frac{2ik}{z^{2}}+\frac{2z-3}{z^{2}}
=2i\mathcal{A}\epsilon+\frac{2+2i\mathcal{B}\epsilon}{z}+\frac{-3+2i\mathcal{C}}{z^{2}}.
\end{equation}
We take the ansatz for $F$ as
\begin{equation}
F=\alpha^{-1} V_{X}=\left(1-\frac{1}{z}\right)
\left(\frac{\lambda +\beta-3i\mathcal{B}\epsilon}{z^{2}}-\frac{3-6i\mathcal{C}}{z^{3}}\right)~.
\end{equation}
The conditions \eqref{C-transf3A} and \eqref{C-transf3B} fix the constants $\alpha$ and $\beta$ as
\begin{equation}
\alpha=\frac{1-\frac{2}{3}i\mathcal{C}}{1-2i\mathcal{C}}~, \quad
\beta=2(1+i\mathcal{B}\epsilon)\alpha^{-1}~.
\end{equation}
$V_{X}$ is obtained as
\begin{equation}
V_{X}=\left(1-\frac{1}{z}\right)
\left(\frac{\alpha\lambda + 2 + (2-3\alpha)i\mathcal{B}\epsilon}{z^{2}}-\frac{3-2i\mathcal{C}}{z^{3}}\right)~.
\end{equation}
Then the resulting differential equation for $X$ is of the form
\begin{equation}
\left[\left(\frac{z-1}{z}\frac{d}{dz}\right)^{2}+p^{2}
-\left(1-\frac{1}{z}\right)\left(\frac{\tilde{\lambda}+2}{z^{2}}
-\frac{3-4i\mathcal{C}}{z^{3}}\right)\right]X=0~,
\label{neweq}
\end{equation}
where $p=k/z^{2}$ and $\tilde{\lambda}$ is defined by
\begin{equation}
\tilde{\lambda}=\alpha\lambda+3i(1-\alpha)\mathcal{B}\epsilon.
\end{equation}
Note that \eqref{neweq} is reduced to the Regge-Wheeler equation under the limit $a \to 0$.

The behavior of the solution for the transformed equation at the boundary and how it is related to that for the Teukolsky equation can also be examined. The solution of the Teukolsky equation \eqref{Tzk} at the boundary behaves as
\begin{equation}
R\sim
\begin{cases}
B_{\mathrm{in}}z^{-1}e^{-i\mathcal{A}\epsilon z^{\ast}}
+B_{\mathrm{out}}z^{3}e^{i\mathcal{A}\epsilon z^{\ast}}\quad \qquad\qquad\textrm{for}\ z\to \infty\ (z^{\ast}\to\infty),
\\
\bar{B}_{\mathrm{in}}z^{2}(z-1)^{2}e^{-i(\epsilon+\mathcal{C})z^{\ast}}
+\bar{B}_{\mathrm{out}}e^{i(\epsilon+\mathcal{C})z^{\ast}}\quad \textrm{for}\ z\to 1 \ (z^{\ast}\to -\infty),
\end{cases}
\end{equation}
where $B_{\mathrm{in}}$, $B_{\mathrm{out}}$, $\bar{B}_{\mathrm{in}}$ and $\bar{B}_{\mathrm{out}}$ are all constant.
On the other hand, the solution of the transformed equation \eqref{neweq} at the boundary behaves as
\begin{equation}
X\sim
\begin{cases}
A_{\mathrm{in}}e^{-i\mathcal{A}\epsilon z^{\ast}}
+A_{\mathrm{out}}e^{i\mathcal{A}\epsilon z^{\ast}}\quad \qquad\textrm{for}\ z\to \infty\ (z^{\ast}\to\infty),
\\
\bar{A}_{\mathrm{in}}e^{-i(\epsilon+\mathcal{C})z^{\ast}}
+\bar{A}_{\mathrm{out}}e^{i(\epsilon+\mathcal{C})z^{\ast}}\quad \textrm{for}\ z\to 1 \ (z^{\ast}\to -\infty),
\end{cases}
\end{equation}
where $A_{\mathrm{in}}$, $A_{\mathrm{out}}$, $\bar{A}_{\mathrm{in}}$ and $\bar{A}_{\mathrm{out}}$ are all constant.
From the Chandrasekhar transformation \eqref{C-transf1} with $Y=z^{-3}R$, the relations between these coefficients
can be found as
\begin{align}
A_{\mathrm{in}}&=-4\mathcal{A}^{2}\epsilon^{2}\alpha \zeta^{-1}B_{\mathrm{in}},
\label{ABAB1}
\\
A_{\mathrm{out}}&=(-4\mathcal{A}^{2}\epsilon^{2})^{-1}\alpha B_{\mathrm{out}},
\label{ABAB2}
\\
\bar{A}_{\mathrm{in}}&=2\alpha \zeta^{-1}[1-2i(\epsilon+\mathcal{C})][1-i(\epsilon+\mathcal{C})]\bar{B}_{\mathrm{in}},
\label{ABAB3}
\\
\bar{A}_{\mathrm{out}}&=\frac{i}{2}(\epsilon+\mathcal{C})^{-1}[1-2i(\epsilon+\mathcal{C})]^{-1}\bar{B}_{\mathrm{out}}.
\label{ABAB4}
\end{align}
Here the constant $\zeta$ is given by
\begin{equation}
\zeta=\alpha^{-1}(\tilde{\lambda}-3i\mathcal{B}\epsilon)(\tilde{\lambda}+2-i\mathcal{B}\epsilon)
-6i\mathcal{A}(1-2i\mathcal{C})\epsilon,
\end{equation}
which is reduced to $(l-1)l(l+1)(l+2)-6i\epsilon$ under the limit $a \to 0$.
The relations \eqref{ABAB1}--\eqref{ABAB4} imply that at each boundary the ``in'' mode and the ``out'' mode
are not mixed under the Chandrasekhar transformation. Then the boundary condition for no energy inflow in the
Teukolsky equation $B_{\mathrm{in}}=\bar{B}_{\mathrm{out}}=0$ is mapped to that in the transformed equation
$A_{\mathrm{in}}=\bar{A}_{\mathrm{out}}=0$ unless the frequency $\epsilon$ is equal to the zeroes or the poles
of the coefficients in \eqref{ABAB1} and \eqref{ABAB4}. Therefore, we can conclude that the
spectra of the quasi-normal modes in the Teukolsky equation and those in the transformed equation coincide generically.

\section{Comparison with quantum Seiberg-Witten geometry}

As well as in the Regge-Wheeler and the Teukolsky equations,
the (confluent) Heun's equation also appears in the quantization of the Seiberg-Witten curves
in supersymmetric gauge theories. For example in $\mathcal{N}=2$ supersymmetric SU(2) gauge theory coupled with
three matter hypermultiplets in the fundamental representation of the gauge group,
the quantum Seiberg-Witten geometry gives the following differential equation
\cite{NS,Gaiotto:2009ma,Aminov:2020yma,Hatsuda:2020iql,Bonelli:2021uvf}
\begin{equation}
\left[\hbar^{2}\frac{d^2}{dz^{2}}+\frac{q(z)}{z^{2}(z-1)^{2}}\right]\Psi(z)=0,
\label{qSW}
\end{equation}
where $\hbar$ is the quantization parameter (hereafter chosen as unity) and $q(z)$ is the quartic polynomial of $z$ as
\begin{align}
q(z)&=\widehat{A}_{0}+\widehat{A}_{1}z+\widehat{A}_{2}z^{2}+\widehat{A}_{3}z^{3}+\widehat{A}_{4}z^{4},
\\
\widehat{A}_{0}&=-\frac{(m_{1}-m_{2})^{2}}{4}+\frac{\hbar^{2}}{4},
\notag\\
\widehat{A}_{1}&=-E-m_{1}m_{2}-\frac{m_{3}\Lambda_{3}}{8}-\frac{\hbar^{2}}{4},
\notag\\
\widehat{A}_{2}&=E+\frac{3m_{3}\Lambda_{3}}{8}-\frac{\Lambda_{3}^{2}}{64}+\frac{\hbar^{2}}{4},
\notag\\
\widehat{A}_{3}&=-\frac{m_{3}\Lambda_{3}}{4}+\frac{\Lambda_{3}^{2}}{32},
\notag\\
\widehat{A}_{4}&=-\frac{\Lambda_{3}^{2}}{64}.
\label{CHEparam}
\end{align}
Here $m_{1}$, $m_{2}$ and $m_{3}$ are the masses of the matter hypermultiplets, $E$ is a moduli parameter
and $\Lambda_{3}$ is the dynamical scale. For generic choice of the parameters, \eqref{qSW} is the form of CHE.
Note that the symmetry under the exchange between $m_{1}$ and $m_{2}$
is manifest because $q(z)$ is unchanged, on the other hand the symmetry under the exchange between $m_{3}$ and another mass
is not manifest\footnote{In the solution with low frequency expansion \cite{Mano:1996vt},
this symmetry becomes manifest \cite{Casals:2021ugr}. }
and the form of the differential equation is changed. However the origin of the parameters suggests that
they describe the same physics, and hence there has to exist some correspondence between them.
By this reason, we will use the above parametrization \eqref{CHEparam} instead of the standard parametrization \cite{Heun}
of CHE. For the relation between these parametrization, see \cite{Bonelli:2021uvf}.

The Regge-Wheeler equation and the Teukolsky equation can be mapped as the form of \eqref{qSW}
by the multiplication transformation. In the case of Schwarzschild spacetime,
the correspondence of the parameters are given by%
\footnote{In the correspondence hereafter, there are three double signs appear in general \cite{Bonelli:2021uvf}.
We have fixed these signs in our convenience.}
\begin{gather}
\hbar=1,\quad \Lambda_{3}=8i\epsilon,\quad E=-l(l+1)+2\epsilon^{2}-\frac{1}{4}~,
\notag\\
m_{1}=-2+i\epsilon,\quad m_{2}=2+i\epsilon, \quad m_{3}=i\epsilon,
\label{SSRWpara}
\end{gather}
for the Regge-Wheeler equation \eqref{RWz} and
\begin{gather}
\hbar=1,\quad \Lambda_{3}=8i\epsilon,\quad E=-l(l+1)+2\epsilon^{2}-\frac{1}{4}~,
\notag\\
m_{1}=-2+i\epsilon,\quad m_{2}=i\epsilon, \quad m_{3}=2+i\epsilon,
\label{SSTpara}
\end{gather}
for the Teukolsky equation \eqref{Tz}. By comparing \eqref{SSRWpara} and \eqref{SSTpara}
one can find that the only difference is the exchange  between $m_{2}$ and $m_{3}$.
And as expected, the solutions to those two equations are related by the Chandrasekhar transformation \eqref{C-transf}.
In the Teukolsky equation in Kerr spacetime, the correspondence of the parameters are computed as
\begin{gather}
\hbar=1,\quad \Lambda_{3}=8i\mathcal{A}\epsilon,\quad E=-\lambda-2+2\epsilon^{2}-\frac{am}{M}\epsilon-\frac{1}{4}~,
\notag\\
m_{1}=-2+i\epsilon,\quad m_{2}=i\epsilon+2i\mathcal{C}, \quad m_{3}=2+i\epsilon.
\label{KTpara}
\end{gather}
On the other hand, the correspondence of the parameters for the differential equation \eqref{neweq} becomes
\begin{gather}
\hbar=1,\quad \Lambda_{3}=8i\mathcal{A}\epsilon,\quad E=-\tilde{\lambda}-2+2\epsilon^{2}-\frac{am}{M}\epsilon-\frac{1}{4}~,
\notag\\
m_{1}=-2+i\epsilon+2i\mathcal{C},\quad m_{2}=2+i\epsilon, \quad m_{3}=i\epsilon.
\label{newpara}
\end{gather}
By comparing \eqref{KTpara} and \eqref{newpara}, one can find that it is not only the exchange between $m_{2}$ and $m_{3}$,
but also $2i\mathcal{C}$ is moved to $m_{1}$, and $\lambda$ is replaced with $\tilde{\lambda}$.
Note that the Regge-Wheeler-type equation which has the parameters in \eqref{KTpara} with the exchange between
$m_{2}$ and $m_{3}$ is given in \cite{Hatsuda:2020iql}.
However, that equation is found just by the exchange of the
parameters, not by the Chandresekhar(-like) transformation.


\section{Summary and discussion}

In this paper, we have proposed the new kind of the Chandrasekhar transformation for the Teukolsky equation
in Kerr spacetime, which reduces to the original work of Chandrasekhar under the limit $a \to 0$.
The original Chandrasekhar transformation had been obtained from the
point view of the gauge transformation for the linearized Einstein
equation. We could expect that our transformation would have a simlilar
origin, and it would be interesting to find it. One can also find that obviously
there could be other transformation by different choices of the variable $x$,
the differential operators $\Lambda_{\pm}$ and
the multiplication transformation and the functions $F$ and $G$.
Well-known examples are of course the (Chandrasekhar-)Detweiler equation and the Sasaki-Nakamura equation
\cite{Detweiler:1977gy,Chandrasekhar:1976zz,Sasaki:1981sx},
where the tortoise coordinate \eqref{tortoise2} is used.
Instead, here we have used the coordinate $z^{\ast}$ defined by \eqref{tortoise1} from \eqref{z}, in order to keep
the structure of the singularities. We have obtained the differential equation \eqref{neweq},
which is also reduced to the Regge-Wheeler equation in the limit $a \to 0$ as
the (Chandrasekhar-)Detweiler and the Sasaki-Nakamura equations are.
The extension to the case of different spins (the scalar waves and the electromagnetic waves)
\cite{Chandrasekhar:1976,Detweiler:1976zz,Nakajima:2020fhi}
would also be interesting.
We have also considered the behavior of the solution for the transformed equation at the boundary
(far infinity and the outer horizon) and have shown that the spectra of the quasi-normal modes in the transformed equation
are generically the same as those in the Teukolsky equation.
The explicit numerical evaluation of the quasi-normal modes and comparison with the results from another method
\cite{Aminov:2020yma,Hatsuda:2020iql,Bonelli:2021uvf,Hatsuda:2020egs,Bianchi:2021xpr,daCunha:2021jkm,Bianchi:2021mft}
would be interesting.

We have also given the interpretation of our transformation using the formalism which is recently
proposed in the study of supersymmetric gauge theories. The Regge-Wheeler equation and the Teukolsky equation
are examples of CHE, 
which also appears as the wave equation for the quantum Seiberg-Witten geometry.
It turns out that our transformation is more non-trivial than the exchange
of the mass parameters.
Then at present there is no explanation why the spectra has to be the same
in the side of the quantum Seiberg-Witten geometry
and we may have to consider more non-trivial transformation as in \cite{Kubo:2018cqw,Moriyama:2021mux}.
A similar analysis for the (Chandrasekhar-)Detweiler equation and
the Sasaki-Nakamura equation would also be useful.

Another possible generalization is to include the cosmological constant.
The Regge-Wheeler and the Teukolsky equations in the background of the Kerr-de Sitter black hole
are examples of the Heun's equation (HE) \cite{Suzuki:1998vy}, which has four regular singularities.
Since HE does not have the irregular singularity, it is slightly easier to handle it than CHE.
The local solution of HE is included in Mathematica, as well as that of CHE.
Some problems about evapolation and scattering are also discussed without approximation
\cite{Gregory:2021ozs,Motohashi:2021zyv}.
These exact analyses would help to study the problem with the arbitrary frequency,
which is important for application of the scattering of the gravitational (electromagnetic or scalar) waves
to more general cases.

\section*{Acknowledgements}
HN thanks to Sung-Soo Kim, Xiaobin Li and in particular Futoshi Yagi for valuable comments and discussions.
This work was supported in part by the National Natural Science Foundation of China (Grant No. 11973025).





\begin{thebibliography}{99}

\bibitem{Abbott:2016blz}
B.~Abbott \textit{et al.} [LIGO Scientific and Virgo],
Phys. Rev. Lett. \textbf{116}, no.6, 061102 (2016)
doi:10.1103/PhysRevLett.116.061102
[arXiv:1602.03837 [gr-qc]].


\bibitem{Mino:1997bx}
Y.~Mino, M.~Sasaki, M.~Shibata, H.~Tagoshi and T.~Tanaka,
Prog. Theor. Phys. Suppl. \textbf{128}, 1-121 (1997)
doi:10.1143/PTPS.128.1
[arXiv:gr-qc/9712057 [gr-qc]].



\bibitem{Regge:1957td}
  T.~Regge and J.~A.~Wheeler,
  Phys.\ Rev.\  {\bf 108}, 1063 (1957)
  doi:10.1103/PhysRev.108.1063.





\bibitem{Teukolsky:1973ha}
  S.~A.~Teukolsky,
  Astrophys.\ J.\  {\bf 185}, 635 (1973)
  doi:10.1086/152444.

\bibitem{Chandrasekhar:1975}
  S.~Chandrasekhar,
   Proc. R. Soc. Lond. A \textbf{A343}, 289-298 (1975)
   doi:10.1098/rspa.1975.0066

\bibitem{Detweiler:1977gy}
S.~L.~Detweiler,
Proc. Roy. Soc. Lond. A \textbf{A352}, 381-395 (1977)
doi:10.1098/rspa.1977.0005

\bibitem{Chandrasekhar:1976zz}
S.~Chandrasekhar and S.~L.~Detweiler,
Proc. Roy. Soc. Lond. A \textbf{A350}, 165-174 (1976)
doi:10.1098/rspa.1976.0101.

\bibitem{Sasaki:1981sx}
M.~Sasaki and T.~Nakamura,
Prog. Theor. Phys. \textbf{67}, 1788 (1982)
doi:10.1143/PTP.67.1788

\bibitem{Darboux}
  G.~Darboux,
  C.R.Academy Sci.(Paris) 94 (1882) 1456
  [arXiv:physics/9908003 [physics]].

\bibitem{Glampedakis:2017rar}
K.~Glampedakis, A.~D.~Johnson and D.~Kennefick,
Phys. Rev. D \textbf{96}, no.2, 024036 (2017)
doi:10.1103/PhysRevD.96.024036
[arXiv:1702.06459 [gr-qc]].

\bibitem{Yurov:2018ynn}
A.~V.~Yurov and V.~A.~Yurov,
Phys. Lett. A \textbf{383}, no.22, 2571-2578 (2019)
doi:10.1016/j.physleta.2019.05.024
[arXiv:1809.10279 [gr-qc]].

\bibitem{Heun}
A.~Ronveaux,
``Heun's Differential Equations,''
Oxford University Press, Oxford, New York, October 1995.

\bibitem{Mano:1996vt}
S.~Mano, H.~Suzuki and E.~Takasugi,
Prog. Theor. Phys. \textbf{95}, 1079-1096 (1996)
doi:10.1143/PTP.95.1079
[arXiv:gr-qc/9603020 [gr-qc]].

\bibitem{Casals:2021ugr}
M.~Casals and R.~T.~da Costa,
[arXiv:2105.13329 [gr-qc]].

\bibitem{Aminov:2020yma}
G.~Aminov, A.~Grassi and Y.~Hatsuda,
[arXiv:2006.06111 [hep-th]].

\bibitem{Hatsuda:2020iql}
Y.~Hatsuda,
[arXiv:2007.07906 [gr-qc]].

\bibitem{Seiberg:1994rs}
N.~Seiberg and E.~Witten,
Nucl. Phys. B \textbf{426}, 19-52 (1994)
[erratum: Nucl. Phys. B \textbf{430}, 485-486 (1994)]
doi:10.1016/0550-3213(94)90124-4
[arXiv:hep-th/9407087 [hep-th]].

\bibitem{Seiberg:1994aj}
N.~Seiberg and E.~Witten,
Nucl. Phys. B \textbf{431}, 484-550 (1994)
doi:10.1016/0550-3213(94)90214-3
[arXiv:hep-th/9408099 [hep-th]].

\bibitem{NS}
N.~A.~Nekrasov and S.~L.~Shatashvili,
``Quantization of integrable systems and four dimensional gauge theories,''
XVIth International Congress on Mathematical Physics,
World Scientific, March 2010, pp. 265-289.

\bibitem{Alday:2009aq}
L.~F.~Alday, D.~Gaiotto and Y.~Tachikawa,
Lett. Math. Phys. \textbf{91}, 167-197 (2010)
doi:10.1007/s11005-010-0369-5
[arXiv:0906.3219 [hep-th]].

\bibitem{Gaiotto:2009ma}
D.~Gaiotto,
J. Phys. Conf. Ser. \textbf{462}, no.1, 012014 (2013)
doi:10.1088/1742-6596/462/1/012014
[arXiv:0908.0307 [hep-th]].

\bibitem{Bonelli:2021uvf}
G.~Bonelli, C.~Iossa, D.~P.~Lichtig and A.~Tanzini,
[arXiv:2105.04483 [hep-th]].

\bibitem{Whiting:1988vc}
B.~F.~Whiting,
J. Math. Phys. \textbf{30}, 1301 (1989)
doi:10.1063/1.528308

\bibitem{Andersson:2016epf}
L.~Andersson, S.~Ma, C.~Paganini and B.~F.~Whiting,
J. Math. Phys. \textbf{58}, no.7, 072501 (2017)
doi:10.1063/1.4991656
[arXiv:1607.02759 [gr-qc]].

\bibitem{Leaver:1986gd}
E.~W.~Leaver,
Phys. Rev. D \textbf{34}, 384-408 (1986)
doi:10.1103/PhysRevD.34.384

\bibitem{Chandrasekhar:1976}
S,~Chandrasekhar,
Proc. R. Soc. Lond. A \textbf{A348}, 39-55 (1976)
doi:10.1098/rspa.1976.0022.

\bibitem{Detweiler:1976zz}
S.~L.~Detweiler,
Proc. Roy. Soc. Lond. A \textbf{A349}, 217-230 (1976)
doi:10.1098/rspa.1976.0069

\bibitem{Nakajima:2020fhi}
H.~Nakajima and W.~Lin,
Class. Quant. Grav. \textbf{38}, no.2, 027001 (2020)
doi:10.1088/1361-6382/abc370

\bibitem{Hatsuda:2020egs}
Y.~Hatsuda and M.~Kimura,
Phys. Rev. D \textbf{102}, no.4, 044032 (2020)
doi:10.1103/PhysRevD.102.044032
[arXiv:2006.15496 [gr-qc]].

\bibitem{Bianchi:2021xpr}
M.~Bianchi, D.~Consoli, A.~Grillo and J.~F.~Morales,
[arXiv:2105.04245 [hep-th]].

\bibitem{daCunha:2021jkm}
B.~C.~da Cunha and J.~P.~Cavalcante,
Phys. Rev. D \textbf{104}, no.8, 084051 (2021)
doi:10.1103/PhysRevD.104.084051
[arXiv:2105.08790 [hep-th]].

\bibitem{Bianchi:2021mft}
M.~Bianchi, D.~Consoli, A.~Grillo and J.~F.~Morales,
[arXiv:2109.09804 [hep-th]].

\bibitem{Kubo:2018cqw}
N.~Kubo, S.~Moriyama and T.~Nosaka,
JHEP \textbf{01}, 210 (2019)
doi:10.1007/JHEP01(2019)210
[arXiv:1811.06048 [hep-th]].

\bibitem{Moriyama:2021mux}
S.~Moriyama and Y.~Yamada,
SIGMA \textbf{17}, 076 (2021)
doi:10.3842/SIGMA.2021.076
[arXiv:2104.06661 [math.QA]].

\bibitem{Suzuki:1998vy}
H.~Suzuki, E.~Takasugi and H.~Umetsu,
Prog. Theor. Phys. \textbf{100}, 491-505 (1998)
doi:10.1143/PTP.100.491
[arXiv:gr-qc/9805064 [gr-qc]].

\bibitem{Gregory:2021ozs}
R.~Gregory, I.~G.~Moss, N.~Oshita and S.~Patrick,
Class. Quant. Grav. \textbf{38}, no.18, 185005 (2021)
doi:10.1088/1361-6382/ac1a68
[arXiv:2103.09862 [gr-qc]].

\bibitem{Motohashi:2021zyv}
H.~Motohashi and S.~Noda,
PTEP \textbf{2021}, 083
doi:10.1093/ptep/ptab097
[arXiv:2103.10802 [gr-qc]].









\end{thebibliography}
\end{document}